\documentclass[onecolumn]{SSA-BSSA}

\citethisauthor{Shcherbakov, R. and S. Kothari}
\doi{00.0000/0000000000}
\recdate{00 Month 2025}

\begin{document}

\title{Earthquake Declustering Using Supervised Machine Learning}

\author[*1,2\orc{0000-0002-3057-0851}]{Robert Shcherbakov}
\author[1]{Sidhanth Kothari}

\affil[1]{Department of Earth Sciences, Western University, London, Ontario, N6A 5B7, Canada}{\auorc[https:// orcid.org/]{0000-0002-3057-0851}{(R.S.)}}
\affil[2]{Department of Physics and Astronomy, Western University, London, Ontario, N6A 3K7, Canada}{}
\corau{*Corresponding author: rshcherb@uwo.ca}

\begin{abstract}
    Earthquake catalog declustering is the procedure of separating event clusters from background seismicity, which is an important task in statistical seismology, earthquake forecasting, and probabilistic seismic hazard analysis. Several declustering methods have been introduced in the literature and operate under the supposition that background events occur independently while clusters are triggered by prior events. Here, we test the ability of Supervised Machine Learning (SML) on the declustering problem by leveraging two popular statistical methods. First, the Epidemic Type Aftershock Sequence (ETAS) model is fit to a target catalog and the parameters are used to generate synthetic earthquake data, which replicate the magnitude-space-time seismicity of the target catalog. Next, the Nearest-Neighbour Distance (NND) metrics are computed between each simulated event and used as features to train the SML algorithm. Finally, the trained algorithm is applied to decluster synthetic testing data and then the original target catalog. Our results indicate that SML method performs better than the NND-based and stochastic declustering methods on the test data and makes more nuanced selections of background and clustered events when applied to real seismicity. While the vast majority of the SML technique's predictive power appears to lie within the NND values of the ''first'' nearest-neighbours, a machine learning analysis reveals that predictive accuracy can be improved by additional ''next'' nearest-neighbours and differential magnitude features. The developed approach is applied to seismic catalogs in Southern California and Italy to decluster them.
\end{abstract}

\maketitle

\begin{keypoints}
\item
A supervise machine learning algorithm is implemented for earthquake declustering.

\item
The algorithm is tested against two other declustering algorithms to demonstrate its better performance.

\item
The machine learning declustering algorithm is applied to seismicity in Southern California and Italy.
\end{keypoints}

\section{Introduction}

Earthquakes form clusters in space and time. Slow tectonic loading redistributes subsurface stresses that leads to the occurrence of random and most likely independent background events with rates varying in space but roughly constant across time. However, background earthquakes can trigger secondary events or \emph{aftershocks}. Aftershocks themselves, in turn, can trigger subsequent events forming hierarchical aftershock sequences. Moreover, other factors, such as fluid migration and pore pressure changes, occurrence of long-distance seismic events, anthropogenic energy related activities can alter the subsurface state of stress and induce earthquake sequences or swarms. This results in overlapping populations of independent \emph{background} and dependent \emph{clustered} earthquakes, which form the bulk of the observed seismicity.

Typically, regional rates of background seismicity vary in space but remain relatively constant in time. For this reason, background events provide valuable information about regional seismogenesis that can be used in probabilistic earthquake forecasting and hazard assessment \citep{FieldMHP2024a,LlenosMSR2024a}. Clustered events, on the other hand, deviate from Poisson-like behaviour by temporarily changing the seismic rate, where events occur in rapid succession due to the transfer and redistribution of stresses from the previous events or other factors. The analysis of clustering aspects of seismicity can offer further insight into local fault interactions, triggering mechanisms, aftershock sequence evolution, to name a few \citep{ConsoleMC2006a,ShcherbakovTR15a,VariniPZ2020a,ZaliapinB2022a}. Therefore, it is of critical importance that earthquake catalogs can be declustered correctly, in order to study the evolution of seismic activity and to assess the associated long-term seismic hazard. 

Discriminating between background events and aftershocks in real earthquake catalogs is a challenging task \citep{ConsoleMC2006a}. In general, it is very difficult to pinpoint specific triggering factors that predominantly contribute to the occurrence of a given earthquake (i.e. tectonic loading, prior seismicity, or others). As a result, the separation of events or \emph{declustering} into background events and aftershocks is typically performed probabilistically \citep{MolchanD92a}. Aftershocks that occur spatio-temporally distant from their parent events are particularly difficult to discriminate. For this reason, a number of declustering approaches have been proposed. Though there is no consensus about the ''best'' approach, available methods generally utilize the same cataloged features such as the event time, location, and magnitude to construct either: (i) space-time windows around suitably large events (\emph{mainshocks}), where smaller events falling within the windows are classified as direct aftershocks \citep{GardnerK1974a,Reasenberg85a}, or (ii) space-time links between event pairs, where clusters are formed by chaining each event to its closest predecessor and labelling the largest (or sometimes the first) event in the cluster as the mainshock \citep{FrohlichD90a,BaiesiP04a,HainzlSB2006a,ZaliapinGKW08a,ZaliapinB13a,ZaliapinB20a}. Type (i) methods, also known as window methods, are usually the most straightforward to apply but sacrifice complexity by attributing all aftershocks directly to the mainshock. Type (ii) methods, known as linking or cluster methods, are more computationally expensive but provide a more realistic picture of inter-event relationships. However, both approaches depend upon certain parameters, such as window lengths and look-ahead interaction times, which can influence estimations of the background rate.

A third type of the earthquake declustering approach, introduced by \cite{ZhuangOV02a,ZhuangOV04a}, is stochastic rather than deterministic in nature. Stochastic Declustering (SD) involves computing background probabilities for events based on the spatial Epidemic Type Aftershock Sequence (ETAS) model \citep{Ogata98a,Ogata2024a}. The probability of a given background event, occurring at a specific time and spatial location, is calculated as the ratio of the estimated background rate and the instantaneous rate. Stochastic realizations of the background and clustering processes can then be produced through statistical random thinning \citep{ZhuangCOC05a,Zhuang06a,StiphoutZM2012a}. Stochastic declustering was also implemented within the framework of cascading aftershocks introduced by \cite{MarsanL08a}. 

Earthquake declustering is extensively used in a Probabilistic Seismic Hazard Analysis (PSHA) \citep{PetersenSPM2020a,LlenosMSR2024a}. It is performed to estimate a time-independent background rate of seismicity that is used to forecast future seismic activity \citep{Cornell1968a,PetersenSPFM2024a}. This is routinely done for most of the seismogenic zones to produce seismic hazard maps that are used in engineering seismology and insurance industry \citep{GerstenbergerMAP20a}. However, recently aftershocks are also being incorporated into the analysis in assessing the earthquake hazard \citep{LlenosM20a,MizrahiNCW2024a}.

Detailed comparisons and performance of several major declustering algorithms and their application to specific seismogenic regions were performed in the past. \cite{PeresanG2018a} conducted a detailed study of earthquake clustering in Northeastern Italy to analyze their spatial and topological structure. \cite{VariniPZ2020a} performed a comparative analysis of the two declustering algorithms, one based on the standard NND approach and the other based on the SD method. They found that these algorithms produced comparable results in delcustering seismicity in Northeastern Italy. However, these two algorithms produced different topological structures for identified event clusters. \cite{MizrahiNW21a} investigated the effect of declustering algorithms on the frequency-magnitude distribution of the background (independent) events in California. They concluded that the considered declustering methods might introduce a bias in the final declustered catalogs. This was further addressed by \cite{TaroniA2021a}, who investigated the effect of the two declustering methods by \cite{GardnerK1974a} and \cite{Reasenberg85a} on the PSHA applied to Italy.

In this paper, we combine the metrics of a popular linking method, the nearest-neighbour distance (NND) method \citep{BaiesiP04a,ZaliapinGKW08a,ZaliapinB13a}, with the ETAS model \citep{Ogata98a,ZhuangOV02a,Ogata2024a}, in order to train a supervised machine learning classifier to identify background seismicity (and complimentarily, aftershocks). Supervised and unsupervised learning algorithms (also known as neural networks) have been deployed extensively in seismological studies over recent years \citep{MousaviB2023a}, largely due to improvements in network monitoring -- leading to increased data variety and availability -- and increases in computer processing power. However, a trend towards overcomplexity in network architecture has been observed, where deep learning (i.e. multi-layered) models are implemented in situations that may benefit from simpler and more interpretable ones \citep{MignanB2019a}. Here, we train a supervised ensemble learner, called a random forest \citep{Breiman2001a,HastieTF2009a}, to discover differences in the spatio-temporal distributions of background and clustered seismicity, given the same limited information utilized in current declustering methods. Our approach is briefly summarized as follows: (i) estimate the space-time ETAS model parameters for a target earthquake catalog by optimizing the log-likelihood function; (ii) simulate training and testing catalogs using these parameters, which replicate the seismicity distributions of the target catalog with known output labels; (iii) compute and compile NND metrics between each synthetic event as feature data; (iv) train the random forest classifier on the computed features and output labels; (v) test the classifier on synthetic test data and apply to the original target catalog to decluster it.

A similar approach was followed by \cite{Aden-AntoniowFS2022a}, who evaluated the effectiveness of machine learning algorithms given the task of declustering earthquake catalogs. Their investigation was conducted in two parts, where, in the first part, they compared declustering results of supervised and unsupervised learning methods on simulated catalogs, which were generated with the ETAS model using a range of randomly selected parameters. The supervised algorithms were trained on a subset of these synthetic catalogs based on nearest-neighbour feature information. In the second part, they used the best-performing model, a random forest, to demonstrate its utility on real regional catalogs and identified several aftershock sequences.

In the following section, we provide background information on the NND analysis, the ETAS model and stochastic declustering, the random forest classifier, and describe our novel earthquake declustering approach (Methods Section). Next, we test our method on synthetic and real catalogs for Southern California and Italy and compare results with the NND-based and stochastic declustering approaches (Results Section). Finally, we discuss our findings and draw several conclusions about the regional seismicity distributions for Southern California and Italy in Discussion and Conclusions Sections.

\section{Methods}\label{methods}

Here we introduce the two statistical methods that form the basis of our declustering approach: the NND and stochastic declustering methods. We then detail our chosen machine learning classifier, the random forest. Finally, we outline the steps of our own declustering algorithm.

\subsection{The NND analysis}\label{NND}

The NND analysis is a powerful tool that is used for declustering catalogs, studying inter-event distance distributions and analyzing earthquake clustering properties \citep{BaiesiP04a,ZaliapinGKW08a}. It was first conceived by \cite{BaiesiP04a} and expanded upon by \cite{ZaliapinB13a,ZaliapinB13b,ZaliapinB16a,ZaliapinB20a}, who showed that frequency histograms of its core metric, rescaled proximity distance $\eta$, form a bimodal distribution for many regional earthquake catalogs, with one mode representing background seismicity and the other reflecting clustered events.

The rescaled proximity distance $\eta_{ij}$ between two earthquakes $i$ and $j$ is defined as
\begin{equation}\label{etaij}
    \eta_{ij}  = \left\{
    \begin{array}{lc}
      t_{ij}\,(r_{ij})^{d_f}\,10^{-b\,m_i}\,, & \quad \textrm{for} \quad t_{ij} > 0\,, \\
      \infty\,,                               & \quad \textrm{for} \quad t_{ij} \leq 0\,, 
    \end{array}
    \right.
\end{equation}
where $t_{ij}=t_j - t_i$ is the inter-event time between event $i$ and event $j$, $r_{ij}$ is inter-event spatial distance and $m_i$ is the magnitude of the first event $i$. Due to large depth uncertainties present in regional catalogs, $r_{ij}$ is usually computed as the arc length between epicenters. $d_f$ and $b$ are catalog parameters, where $d_f$ is the fractal dimension of spatial distribution of epicenters or hypocenters (typically $1.2 < d_f < 1.6$ for the two dimensional space) and $b$ is the Gutenberg-Richter $b$-value \citep{ZaliapinB13a}.

The proximity distance $\eta_{ij}$ can be decomposed into its rescaled temporal and spatial components as follows: $\eta_{ij} = T_{ij} \cdot R_{ij}$, where
\begin{eqnarray}
    T_{ij} & = & t_{ij}\,10^{-\frac{1}{2}b\,m_i}, \\
    R_{ij} & = & (r_{ij})^{d_f}\,10^{-\frac{1}{2}b\,m_i}\,.
\end{eqnarray}
In the NND analysis, $\eta_{ij}$ is computed for all possible earthquake pairs $(i,j)$ and the minimized quantity $\eta^{(1)}_j = \min\limits_{i:i<j}\eta_{ij}$ is assigned to each event $j$. The event $i$ corresponding to this shortest distance is labeled as the nearest-neighbour and potential parent of event $j$. It is also possible to define the next nearest-neighbor $\eta^{(2)}_j$ and higher order $\eta^{(n)}_j$ rescaled distances by considering the second lowest, third and higher-order minima values of $\eta_{ij}$ between the pairs.

Once the minimum inter-event distances have been determined for each event, frequency histograms for $\eta$ and its components $(T,\,R)$ are plotted. For regional and worldwide seismicity a signal in the distributions at large $\eta$ corresponds to independent activity, while one centered at small $\eta$ indicates spatiotemporal clustering. In the standard NND analysis, a linear threshold $\eta_0$ is applied to separate these two subpopulations, chosen either by inspection or as the intersection of a two-component Gaussian mixture model \citep{ZaliapinB13a,ZaliapinB16a,KothariSA20a,SedghizadehBS2023a}. 

This threshold approach to splitting the distribution is straightforward, convenient and provides a reasonable distinction between background and clustered events. However, it fails to account for the overlap often observed between them \cite{ZaliapinB20a} and therefore misclassifies background events (aftershocks)  that occur spatiotemporally close to (far from) their nearest-neighbours. As a result \cite{ZaliapinB20a} introduced a refined procedure based on the NND method to decluster the catalogs.

\subsection{The spatio-temporal ETAS model}\label{ETAS}

The ETAS model is a self-exciting point process that represents occurrence rates (or ''intensities'') as the sum of a constant background forcing with a clustering productivity term \citep{Ogata98a,Ogata2024a}. Further, the intensity is conditioned by all prior event occurrences, i.e. the seismic history $\mathcal{H}_t$, and can be written as follows \citep{ZhuangOV02a,ZhuangCOC05a}:
\begin{equation}\label{ETASEq}
    \lambda(t, x, y\,|\, \mathcal{H}_t) = \mu\,u(x, y) + \sum_{i:t_i<t}\xi(t-t_i, x-x_i, y-y_i; m_i)\,,
\end{equation}
where the forcing term $\mu\,u(x,y)$ denotes the spatially-inhomogeneous background rate of earthquakes and the clustering productivity $\xi(t, x, y; m)$ symbolizes the contribution of an event of magnitude $m_i$ to the overall rate at $(t, x, y)$.
The clustering productivity $\xi(t, x, y; m)$ can be written in a factorized form \citep{ZhuangCOC05a}:
\begin{equation}\label{xiEq}
    \xi(t, x, y; m) = \kappa(m)\,g(t)\,f(x, y, m),
\end{equation}
where $\kappa(m)$ specifies the productivity of an event of magnitude $m$ as the expected number of triggered events above a reference magnitude $m_0$: 
\begin{equation}\label{kappaEq}
    \kappa(m) = A\,e^{\alpha(m - m_0)}, \quad m \geq m_0\,,
\end{equation}
where $\alpha$ controls the degree of productivity of each event. The temporal kernel $g(t)$ defines how fast the rate decays after each event and is assumed to follow the Omori-Utsu functional form:
\begin{equation}\label{OmUtEq}
    g(t) = \frac{p - 1}{c} \left(1 + \frac{t}{c} \right)^{-p}, \quad t \ge 0\,.
\end{equation}
The spatial kernel $f(x,y,m)$ assumes a hyperbolic rate decay away from the epicenter of an event of magnitude $m$ \citep{OgataZ2006a}:
\begin{equation}\label{spatEq}
    f(x, y, m) = \frac{q - 1}{\pi D^2 e^{\gamma(m - m_0)}} \left[1 + \frac{x^2 + y^2}{D^2 e^{\gamma(m - m_0)}}\right]^{-q}\,,
\end{equation}
where $\gamma$ scales the spatial decay with magnitude. The set of parameters $\hat{\theta} = [\hat{\mu},\hat{A},\hat{\alpha},\hat{c},\hat{p},\hat{D},\hat{q},\hat{\gamma}]$ is estimated from a catalog of a suitable size by iteratively maximizing the log-likelihood function:
\begin{equation}\label{ETASLL}
    \log L(\theta) = \sum_{i:(t_i,x_i,y_i)\in [T_s,T_e]\times S}\log \lambda_\theta(t_i,x_i,y_i \,|\, \mathcal{H}_{t_i}) - \int_{T_s}^{T_e}\iint_{\,S}\log \lambda_\theta(t,x,y \,|\, \mathcal{H}_t)\,dx\,dy\,dt\,,
\end{equation}
where the summation is taken over all events that fall within the target time interval $[T_s, \, T_e]$ and study region $S$.

\subsection{The Stochastic Declustering (SD) method}\label{SD}

Given a conditional intensity function, equation~(\ref{ETASEq}), one can approximate the probability that an event $j$ was triggered by a prior event $i$ by computing the individual contribution of event $i$ relative to the overall intensity at the time and location $(t_j, x_j, y_j)$ of the event $j$ \citep{ZhuangOV02a,ZhuangOV04a,ZhuangCOC05a,Zhuang06a}:
\begin{equation}\label{SDrho}
    \rho_{ij} = \frac{\kappa(m_i)\,g(t_j - t_i)f(x_j - x_i,y_j - y_i; m_i)}{\lambda(t_j,x_j,y_j|\mathcal{H}_{t_j})}\,.
\end{equation}
Likewise, the probability that event $j$ arose independently can be estimated from the relative contribution of the background rate $\mu\,u(x_j,y_j)$ at $(x_j, y_j)$ \citep{ZhuangCOC05a}:
\begin{equation}\label{SDBkgrdProb}
    \phi_j = \frac{\mu\,u(x_j,y_j)}{\lambda(t_j, x_j, y_j|\mathcal{H}_{t_j})}\,.
\end{equation}
Finally, the probability that $j$ was a triggered event is
\begin{equation}\label{SDAfshkProb}
    \rho_j = 1 - \phi_j = \sum_{i<j}\rho_{ij}.
\end{equation}
From the set of $\phi_j$ or $\rho_j$ values, background and clustering subprocesses can be composed through random thinning \citep{ZhuangOV02a,ZhuangCOC05a}. In the SD method, after fitting the ETAS model and estimating the background probabilities $\phi_j$ for each event $j$, random numbers $U_j$ are generated from a uniform distribution in the interval $[0,\,1]$ and each $U_j$ is compared to $\phi_j$. If $U_j < \phi_j$ then event $j$ is labeled as a background event, otherwise it is labeled as a clustered event. In this work, we use stochastic declustering as a benchmark (specifically, one that does not make use of NND methodology) to compare classification accuracies with those of the supervised machine learning and NND approaches.

\subsection{The Random Forest (RF) algorithm}\label{RF}

Decision trees are among several widely-implemented supervised learning methods and are capable of handling classification (discrete output) and regression (continuous output) problems \citep{Breiman2001a,HastieTF2009a}. They learn from a labeled training set by mapping input-to-output values through a series of binary decisions, or splits, based upon specific feature information. A decision tree can be fully described in graphical space, where nodes represent decision splits, directed branches supply logical conjunctions between nodes, and leaves (end-nodes) form the predicted outputs \citep{HastieTF2009a}.

To construct a decision tree, feature thresholds are tested iteratively at each potential split and are evaluated for optimal purity, which corresponds to the minimization of a cost function. For classification trees, the cost may be estimated using the Gini index or Impurity function \citep{JamesWHT2021a}:
\begin{equation}
    I_G(p) = \sum_{n=1}^{K}p_{n}(1-p_n)\,,
\end{equation}
where $K$ is the total number of classes and $p_{n}$ is the probability that an element is of class $n$. For binary classification problems, such as earthquake declustering, the Gini index is bounded by $0 \leq I_{G}(p) \leq 0.5$. An index of $0$ indicates pure splitting of the data into a single class while an index of $0.5$ indicates that the feature threshold in question performs no better than random selection. Probabilities are weighted against their class proportion as classes are often unequally distributed within a dataset. The main idea is to construct the smallest possible tree such that the root employs a threshold that maximizes impurity reduction, while subsequent nodes provide finer reductions and eventually terminate at leaves with minimal possible impurity.

Decision trees have the ability to achieve high prediction accuracy while maintaining interpretability and require relatively little data preprocessing or parameter tuning \citep{HastieTF2009a,JamesWHT2021a}. However, they are also prone to over-fitting -- particularly if allowed to grow overly-complex or introduce splits on small sample sizes \citep{HastieTF2009a}. An effective way to combat overfitting is ensemble learning, which grows an ensemble, or forest, of trees trained on subsamples of both training data and feature sets. Growing many trees (termed ''weak learners'') on bootstrapped data helps reduce model variance while training on random subsets of feature information prevents excessive correlation between learners.

\subsection{The SML declustering algorithm}\label{SML}

Here, we outline the steps of our supervised machine learning declustering algorithm, which builds upon the methods and techniques described above.

\begin{enumerate}
  \item[\textbf{1:}] Estimate the ETAS parameters for a target catalog, based on the conditional intensity given in equation~(\ref{ETASEq}), by iteratively maximizing the log-likelihood function, equation~(\ref{ETASLL});
  \item[\textbf{2:}] Generate labeled synthetic catalogs for training and testing using the parameters estimated in Step \textbf{1}. These catalogs replicate the energy-space-time distributions of the target seismicity within the framework of the ETAS model, using identical conditions for the time interval, study region, magnitude distribution (i.e. $b$-value), and minimum magnitudes;
  \item[\textbf{3:}] Compute NND values \{$\eta^{(1)}$, $T^{(1)}$, $R^{(1)}$\} for the target catalog and for each synthetic catalog. For additional feature information, compute values between the second through the $n$th next nearest-neighbours \{$\eta^{(n)}$, $T^{(n)}$, $R^{(n)}$, $\delta m$\}, where $n=2,\ldots,10$ is the order of the next nearest-neighbor distances, as well as the magnitude differences ($\delta {m}_{ij} = m_i - m_j$) between the first nearest-neighbours. The NND values will form the bulk of training feature information for the random forest classifier. That is, the classifier will use the training data to learn how background events and aftershocks are distributed in relation to their nearest-neighbours;
  \item[\textbf{4:}] Train a random forest algorithm on the features obtained in Step \textbf{3}. Tune hyperparameters (internal learning parameters) using $k$-fold cross-validation. For ensembles, hyperparameters include the number of trees in the forest, the maximum number of splits per tree, the minimum leaf size, and the learning rate. Cross validation also helps to control overfitting and improves model generalization by training and testing on the entire training set;
  \item[\textbf{5:}] Use the trained classifier to decluster synthetic test catalogs and the original target catalog of earthquakes. Evaluate the procured background event distributions in light of a stationary Poisson process.
\end{enumerate}

The numerical implementation of the SML declustering method, ETAS model fitting and forward simulation, NND analysis was done in Matlab software. For the random forest algorithm we used the classifier implemented in Statistics and Machine Learning Toolbox of the Matlab software package.

\subsection{Statistical tests for independence and stationarity in time}

Statistical tests can be applied to a declustered catalog to check whether the declustered events are stationary and independent. Therefore, one can test the \emph{null hypothesis} that the declustered events are the realization of a homogeneous Poisson process \citep{LuenS2012a,ZaliapinB20a}. Here we consider and apply the Kolmogorov-Smirnov (KS) test \citep{LuenS2012a} and the Brown-Zhao (BZ) test \citep{BrownZ2002a,LuenS2012a}.

The Kolmogorov-Smirnov test compares the empirical cumulative distribution of the transformed times of declustered events with the distribution of rescaled times drawn from the stationary Poisson process \citep{LuenS2012a}. The rescaled times for the declustered events are computed as follows:
\begin{equation}\label{rtimes}
    u_i=\frac{t_i-t_\mathrm{min}}{t_\mathrm{max}-t_\mathrm{min}}\,,
\end{equation}
where $t_\mathrm{min} = \min\limits_{i}(t_i)$ and $t_\mathrm{max}=\max\limits_{i}(t_i)$. If the rescaled times follow Poisson statistics then they must be uniformly distributed on the interval $[0,\,1]$. Therefore, one can apply the standard KS test to the cumulative number of the observed rescaled background event times to check whether they follow the Poisson statistics or not. The test also computes the corresponding $p$-value. Typically, the values of $p$ less than a given significance level (typically taken equal to 0.05) signify that the null hypothesis can be rejected.

In case of the Brown-Zhao test, the observational time interval is partitioned into $K$ nonoverlapping time segments of equal length. For each segment with $N_k$, $k=1,\ldots,K$, number of events, one can compute the following quantities: $Y_k=\sqrt{N_k+\frac{3}{8}}$ and $\bar{Y}=\frac{1}{K}\sum_{k=1}^{K}Y_k$. This is used to compute the test statistic \citep{LuenS2012a,ZaliapinB20a}:
\begin{equation}\label{BZ_ts}
    \chi^2_\mathrm{BZ} = 4\sum_{k=1}^{K}\left(Y_k-\bar{Y}\right)^2\,.
\end{equation}
For point processes that follow homogeneous Poisson statistics, the test statistic (\ref{BZ_ts}) is distributed according to the chi-square distribution with $K-1$ degrees of freedom. As a result, the statistical $p$-value can be computed as the probability that the test statistic is equal to or more extreme than the observed value (\ref{BZ_ts}).

\section{Results}\label{results}

In this section, we report on the ability of the SML algorithm introduced above to decluster synthetic and real earthquake catalogs. We compare the obtained results with those derived using the declustering method based on the NND analysis \citep{ZaliapinB20a} and the SD method \citep{ZhuangOV02a,ZhuangCOC05a}. We also perform the statistical tests on the declustered background events to check the hypothesis that the background events form a stationary Poisson process.

\subsection{Declustering the earthquake catalogs for Southern California and Italy}

\begin{figure}[!ht]
\centering
{\includegraphics*[width=.67\textwidth, trim=11mm 56mm 10mm 53mm]{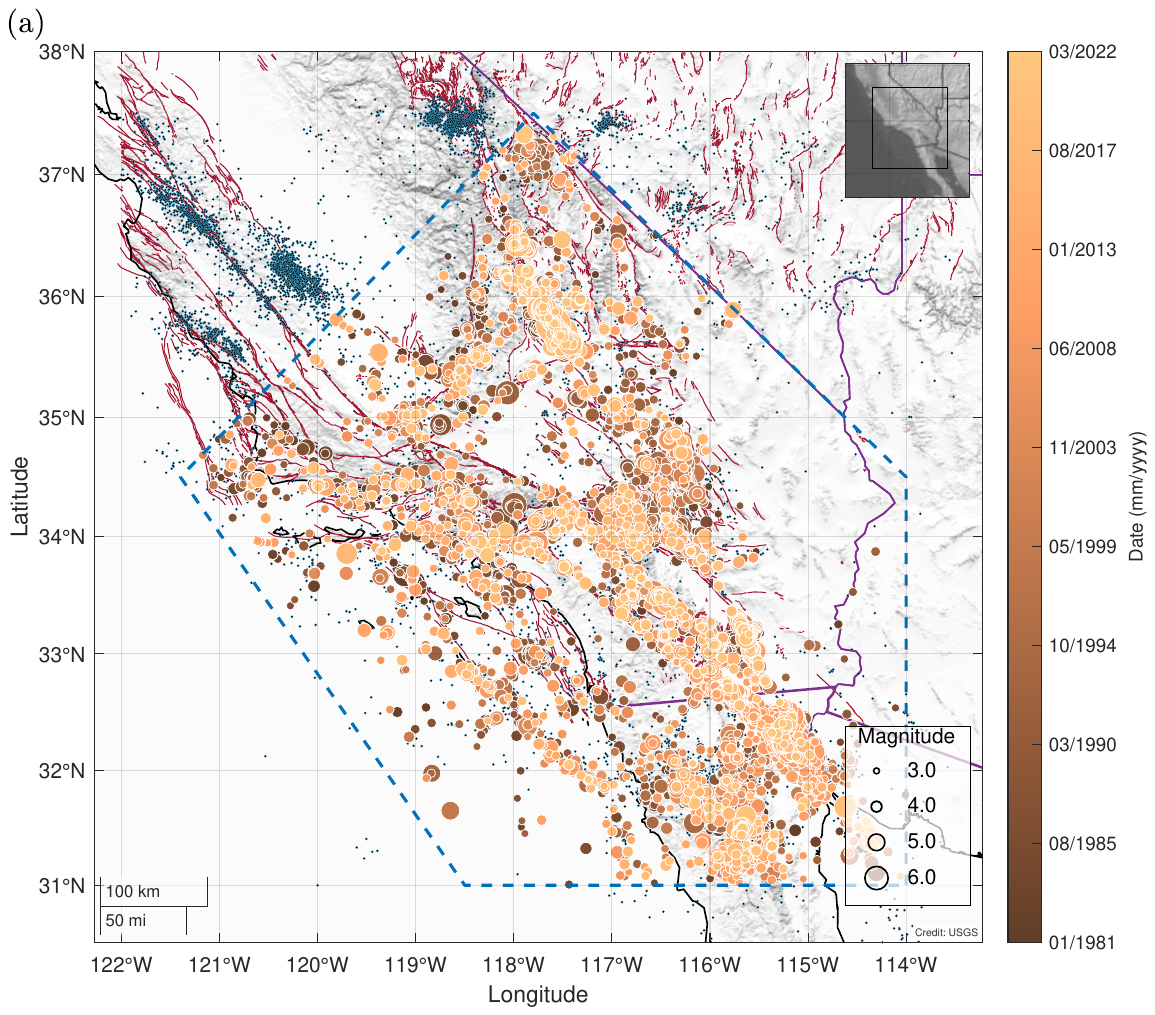}}

{\includegraphics*[width=.67\textwidth, trim=11mm 56mm 10mm 53mm]{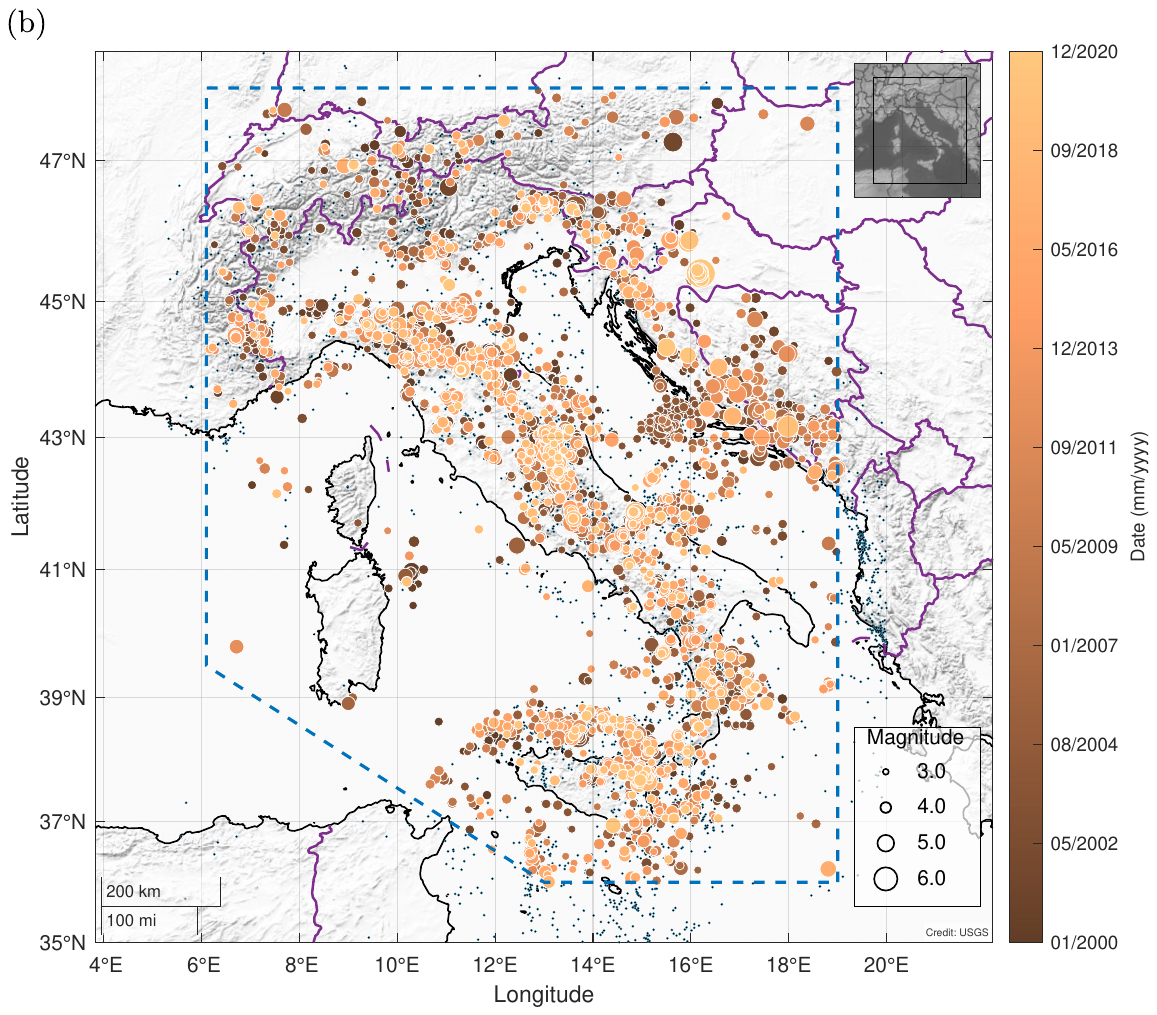}}
\caption{Seismicity maps for (a) Southern California from 1981/01/01 to 2022/03/31 and (b) Italy from 2000/01/01 to 2021/01/01. The colored solid circles are events above magnitude $m\ge 3.0$. Black dots are events below magnitude $m<3.0$. Dashed blue lines outline the target regions.}
\label{fig1}
\end{figure}

We used the regional earthquake catalogs for Southern California \citep{HaukssonYS12a} and Italy (downloaded from Istituto Nazionale di Geofisica e Vulcanologia (INGV) the Italian Seismological Instrumental and Parametric Data-Base (ISIDe) \citep{ISIDe}). For each catalog, the target time interval is defined by its start and end times $[T_s,\,T_e]$, while the preparatory period between initialization and start time $[T_0,\,T_s[$ constitutes the seismic history to stabilize the rate during the target time interval (Table \ref{tab1}). The target region for Southern California is delimited by the polygon illustrated in Figure~\ref{fig1}(a). We use the regional catalog for Italy, bounded by the polygon given in Figure~\ref{fig1}(b). The events between $[T_0,\,T_s[$ and outside the target regions are used to condition the modeled earthquake rate. The ETAS parameters are estimated during the target time interval $[T_s,\,T_e]$ and within the target regions themselves. Other initial parameters are the estimated Gutenberg-Richter $b$-values and the reference magnitude threshold $m_0$ (Table~\ref{tab1}). 

The estimation of the ETAS model parameters was performed using software scripts written in Matlab based on the iterative algorithm outlined in \cite{ZhuangOV02a}. The initial background rate $u(x,y)$ was estimated from the whole catalog and then adjusted after each iteration step by minimizing the log-likelihood function (\ref{ETASLL}). The estimated parameters and the background rate were used to perform the stochastic declustering by using equations (\ref{SDrho})-(\ref{SDAfshkProb}) combined with random thinning \citep{ZhuangOV02a}. The estimated ETAS parameters are reported in Table~\ref{tab2} for Southern California and Italy seismicity. As a result of SD analysis, each event in the real catalogs was classified as a background event or an aftershock.

\begin{table}
\tbl{Seismicity parameters for Southern California and Italy.\label{tab1}}
{\begin{tabular*}{0.75\columnwidth}{@{\extracolsep\fill}lcccccccc}
\toprule
   \textbf{Region}              & $N$     & $T_0$        & $T_s$        & $T_e$        & $b$             & $m_0$  \trowsep
\toprule
   \textbf{Southern California} & $10429$ & $1981/01/01$ & $1991/01/01$ & $2022/03/31$ & $1.04 \pm 0.02$ & $3.0$  \trowsep
   \textbf{Italy}               & $4424$  & $2000/01/01$ & $2004/01/01$ & $2021/11/01$ & $1.17 \pm 0.04$ & $3.0$  \trowsep
\hline\hline
\end{tabular*}}
{}
\end{table}
 
\begin{figure}[!ht]
\centering
{\includegraphics*[width=.46\textwidth, trim=31mm 34mm 28mm 35mm]{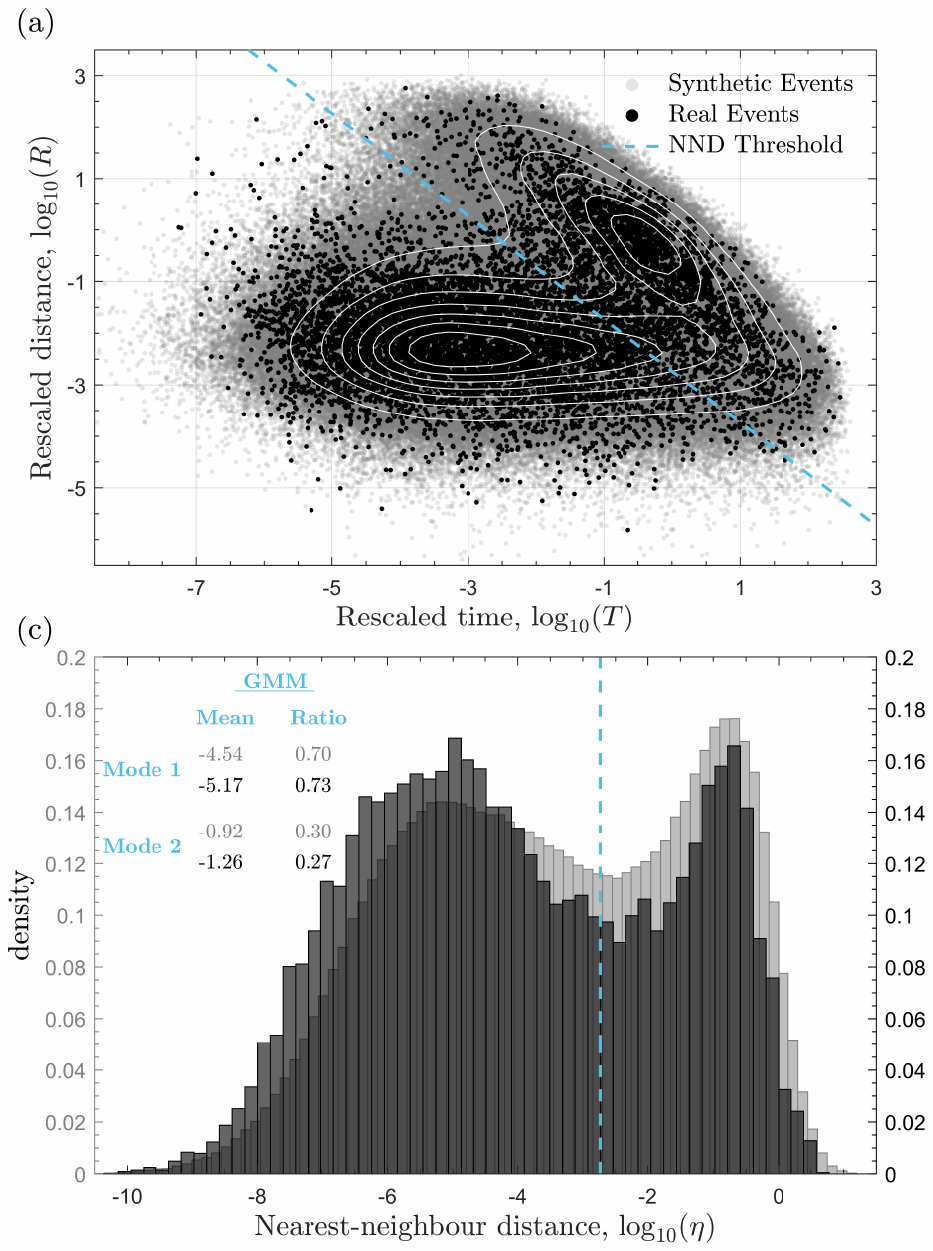}}
\enspace\,
{\includegraphics*[width=.46\textwidth, trim=31mm 34mm 28mm 35mm]{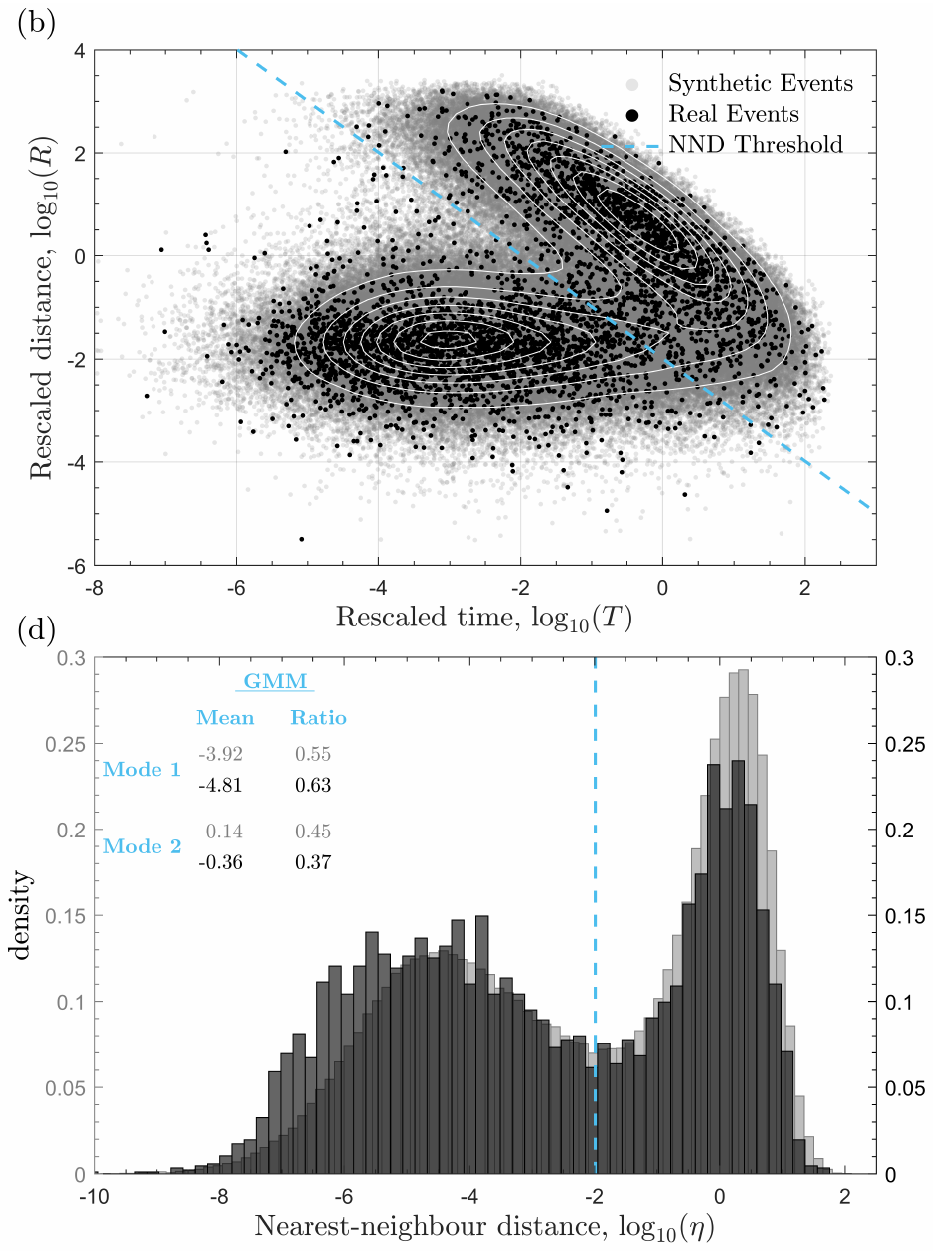}}
\caption{Joint ($T,R$) distributions of nearest-neighbour distances for (a) Southern California and (b) Italy. Dashed blue lines are the estimated thresholds $\log_{10}(\eta_0)$ using the value $\log_{10}(\eta_0)=-2.42$ for Southern California (a,c) and $\log_{10}(\eta_0)=-1.99$ for Italy (b,d). One-dimensional distributions of the proximity distances ($\eta$) for (c) Southern California and (d) Italy.}
\label{fig2}
\end{figure}

We used the estimated ETAS model parameters within the specified target time intervals and spatial regions for Southern California and Italy (Table~\ref{tab2}) to produce synthetic earthquake catalogs. We generated $200$ synthetic catalogs for each region with assigned output (background/aftershock) labels for each event, which were used to train and test the SML algorithm. For the synthetic simulations, we implemented a branching algorithm to simulate the spatio-temporal ETAS model, as a result each event was labeled either a background event or an aftershock. The synthetic events were first simulated during the whole time interval $[T_0,\,T_e]$ but only events in the target time interval $[T_s,\,T_e]$ were retained. The NND distributions for both the real and synthetic catalogs are plotted in Figure~\ref{fig2}. Upon visual inspection, the rescaled distance distributions of the synthetic catalogs appear to match those of the real data (black markers), indicating that the ETAS model is doing a good job of reconstructing the inhomogeneous earthquake rates. This observation is supported by the arrangement of inter-event distances in Figure~\ref{fig2}(a,b), where the space-time-magnitude distributions resemble each other closely (i.e. substantial overlap in scattering and modality between grey and black markers). We applied a two-component Gaussian mixture model to the $\eta$ distributions, Figure~\ref{fig2}(c,d), to compare the mixing proportions and mean locations of the background and aftershock subpopulations. A critical threshold $\eta_0$ was computed as the minimum of a saddle point between the two modes of the Gaussian mixture model. We obtained the following values: $\log_{10}(\eta_0)=-2.42$ for Southern California and $\log_{10}(\eta_0)=-1.99$ for Italy.

\begin{table}
\tbl{The estimated ETAS parameters for Southern California and Italy.\label{tab2}}
{\begin{tabular*}{1.0\columnwidth}{@{\extracolsep\fill}lcccccccc}%
\toprule
   \textbf{Region} & $\mu$ & $A$ & $\alpha$ & $c$ & $p$ & $d$ & $q$ & $\gamma$ \trowsep
\toprule
   \textbf{Southern California} & $0.86 \pm 0.11$ & $0.55 \pm 0.17$ & $1.23 \pm 0.07$ & $0.0035 \pm 0.0017$ & $1.08 \pm 0.03$ & $0.20 \pm 0.01$ & $1.46 \pm 0.04$ & $1.53 \pm 0.15$ \trowsep
   \textbf{Italy} & $0.86 \pm 0.06$ & $0.32 \pm 0.05$ & $1.50 \pm 0.10$ & $0.009 \pm 0.002$ & $1.16 \pm 0.02$ & $1.01 \pm 0.02$ & $1.85 \pm 0.04$ & $1.02 \pm 0.02$ \trowsep
\hline\hline
\end{tabular*}}
{}
\end{table}

Comparing the $\eta$ distributions for two regions, Southern California catalogs (both real and synthetic) contain higher proportions of triggered events (smaller inter-event distances), while Italian seismicity have a more balanced mixture of independent and triggered events. Southern California seismicity also has slightly more overlap between its two modes; they are situated in closer proximity and blend together without an obvious distinction. Conversely, Italy's seismicity two modes are more distinct and better separated. We therefore anticipated the Southern California catalogs would be more difficult to decluster than Italy catalogs. Perhaps another reasonable assumption is that it would be harder to improve upon the NND-based declustering approach for Italy's catalogs, which appear to be nicely partitioned by the linear threshold compared to Southern California catalogs. We show below that both the supervised machine learning and stochastic declustering methods are still able to increase prediction accuracy for both regions.

After generating labeled training catalogs, we computed the NND features and differential magnitude predictor information $\{\eta^{(n)}$, $T^{(n)}$, $R^{(n)}$, $\delta m$\}, where $n=1,\ldots,10$ is the order of the nearest-neighbor distances. The predictors were compiled in tabular form (analogous to the catalogs themselves), where rows corresponded to individual events and columns contained specific features. It is worth noting that the features we chose chiefly provide \emph{relative} event information -- that is, they are indicative of an earthquake's relationship to a prior earthquake. We discarded absolute occurrence times and locations; however, we elected to retain absolute magnitudes on the grounds of their constructed relationship to clustering productivity proposed by the ETAS model. We also included the background probabilities for each event computed during the fitting of the ETAS model. We then iteratively trained ensemble classifiers on $100$ randomly selected synthetic catalogs and tested them on the remaining $100$, in a manner similar to the $k$-fold cross validation technique introduced in the previous section. This allowed us to calculate average scores and reduce the uncertainty in the prediction accuracy.

\begin{figure}[!ht]
\centering
{\includegraphics*[width=.75\textwidth, trim=25mm 3mm 25mm 5mm]{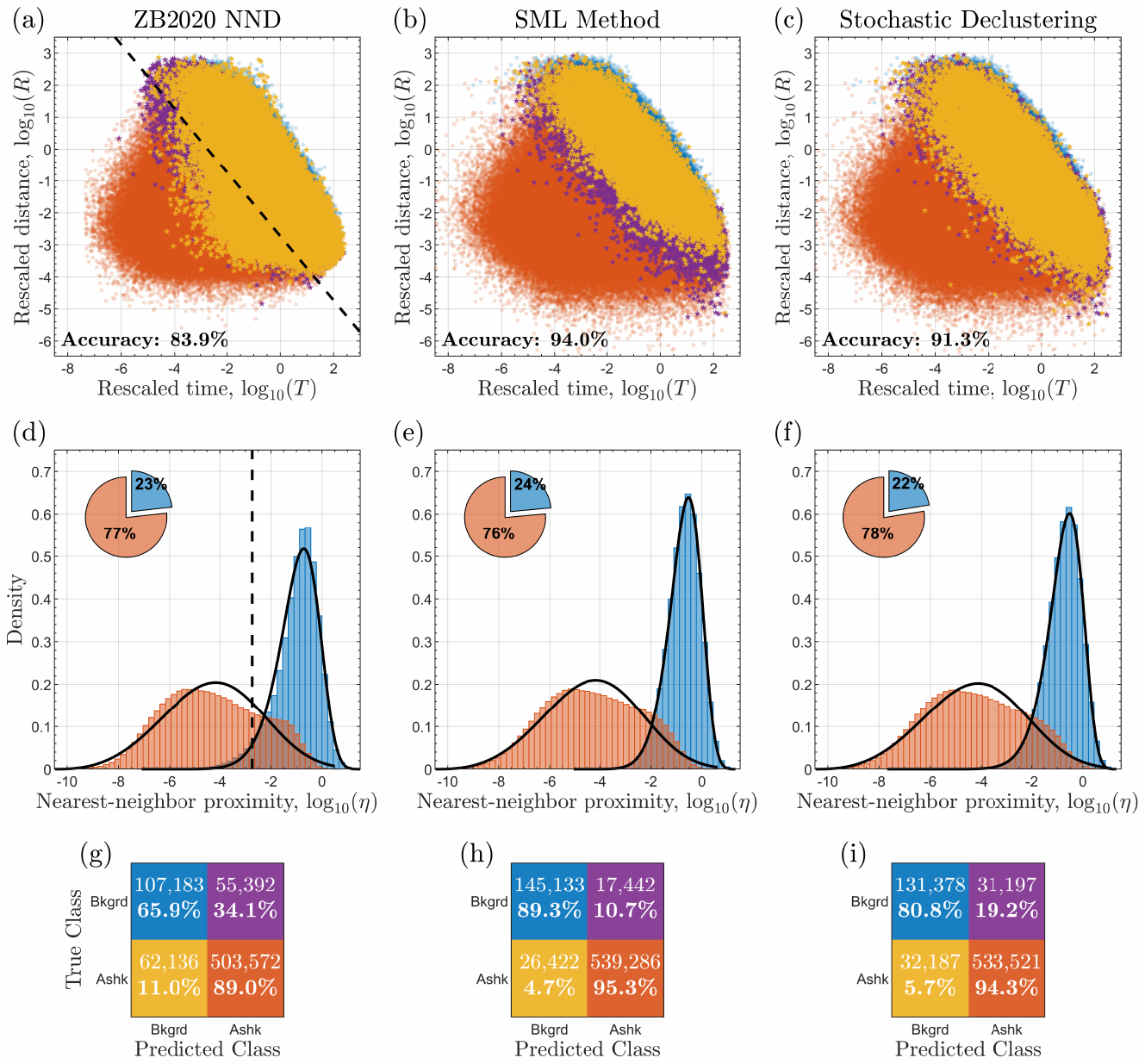}}

\caption{Declustering comparison of the ETAS-generated synthetic testing catalogs for Southern California. Top row (a)-(c): joint ($T,R$) distributions of nearest-neighbour distances. Dashed black lines are the estimated single-value thresholds separating background and clustered modes. The thresholds have no bearing on the SML or stochastic declustering methods. Blue and orange markers are the correctly identified background and aftershock events, respectively. Purple and yellow markers indicate misclassified background and aftershock events. Middle row (d)-(f): normalized one-dimensional ($\eta$) distributions of nearest-neighbour distances. Black curves are Weibull probability density functions. Piecharts provide estimated population percentages. Bottom row (g)-(i): confusion charts for each declustering method.}
\label{fig3}
\end{figure}

The results of the application of the three methods to decluster the synthetic seismicity for Southern California are illustrated in Figure~\ref{fig3}. Similar results obtained by declustering the synthetic seismicity for Italy are reported in Figure~S3. The method of \cite{ZaliapinB20a} requires an additional parameter $\alpha_0$ to perform the declustering. For Southern California we used $\alpha_0=0.7$ and for Italy we used $\alpha_0=1.2$. \cite{ZaliapinB20a} do not provide a specific recommendations for choosing this parameter except stating that it can be in the range $\alpha_0\in [-1,\,1]$. However, this parameter significantly affects the declustering and introduces ambiguity in obtained results. By adjusting this parameter, one can shift the proportion between clustered and background events.

Figure \ref{fig3}(a)-(c) provides NND distributions based on $100$ synthetic catalogs. Marker colors in the top row reflect model predictions (background/aftershock) and results (correct/incorrect). Figure \ref{fig3}(g)-(i) gives confusion charts for each method, which provide a breakdown of absolute and relative prediction results. From the charts, we can see that synthetic testing catalogs were declustered well by all three methods. Southern California contains a subset of aftershock events at larger inter-event distances resulting in a second hump near the intersection of the two modes. This subpopulation is better accounted for by the SML method. For Southern California synthetic seismicity, the SML method provides overall the best performance (94.0\% overall accuracy) compared to the NND-based method (83.8\%) and the SD method (91.3\%). Similarly for Italy, the SML achieves a 4.3\% gain over the NND method and a 1.8\% gain over SD method (Figure~S3). The confusion charts indicate that the SML method correctly identifies more aftershocks and background events. The NND-based method (ZB2020) from \cite{ZaliapinB20a} has the lowest performance accuracy among the three methods.

\begin{figure}[!ht]
\centering
{\includegraphics*[width=.75\textwidth, trim=25mm 27mm 25mm 27mm]{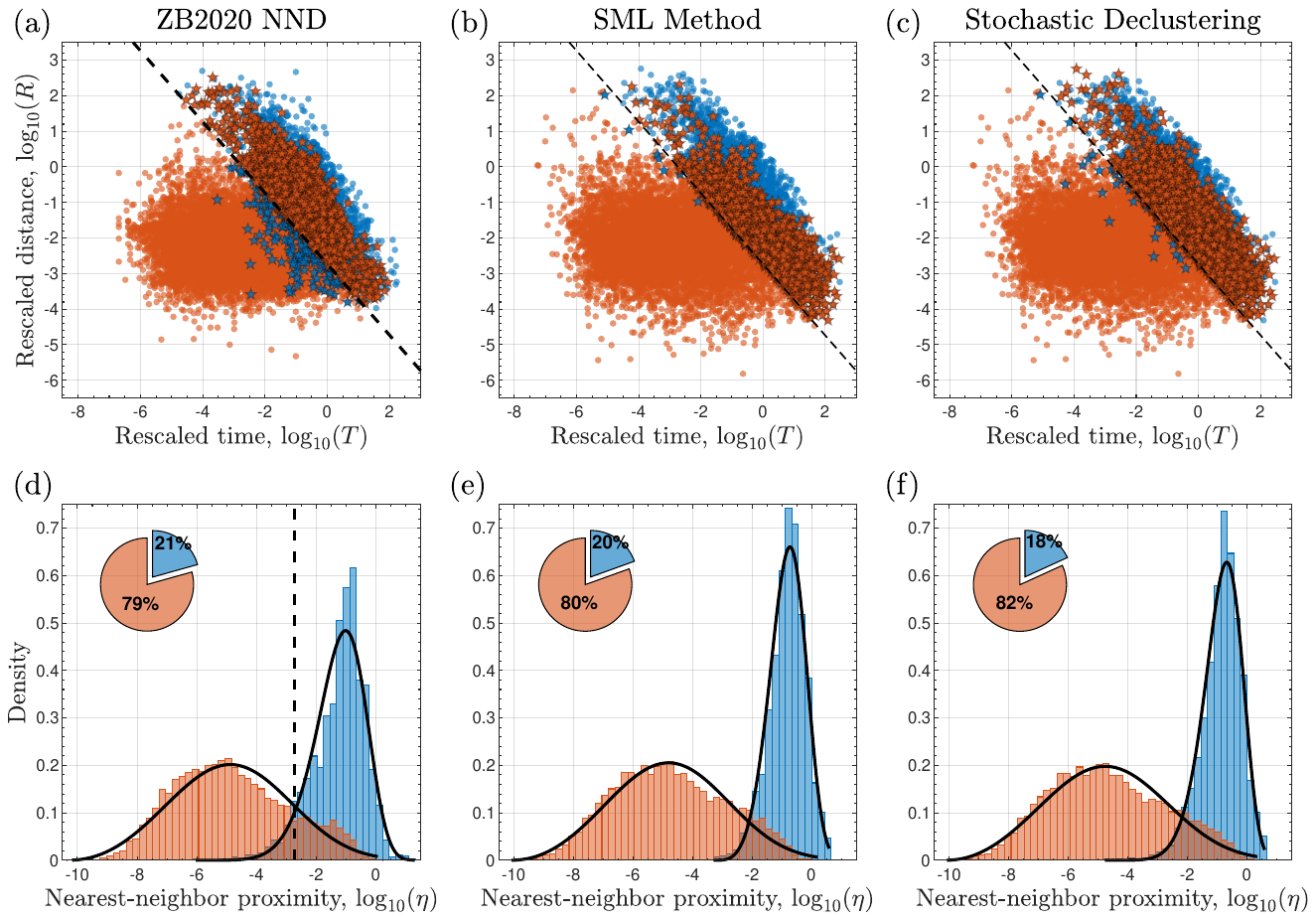}}

\caption{Declustering comparison for the real catalog of Southern California. Top row (a)-(c): joint ($T,R$) distributions of nearest-neighbour distances. Dashed black lines are the estimated single-value thresholds separating background and clustered modes for the original NND method. Blue and orange solid circle markers are the classified background and aftershock events, respectively. Blue and orange stars indicate background and aftershock events classified by the three methods that lie beyond the NND threshold. Bottom row (d)-(f): normalized one-dimensional ($\eta$) distributions of nearest-neighbour distances. Black curves are the fits of the Weibull probability density functions to the distribution of aftershocks and background events. Piecharts provide estimated population percentages.}
\label{fig4}
\end{figure}

\begin{figure}[!ht]
\centering
{\includegraphics*[width=.75\textwidth, trim=25mm 27mm 25mm 27mm]{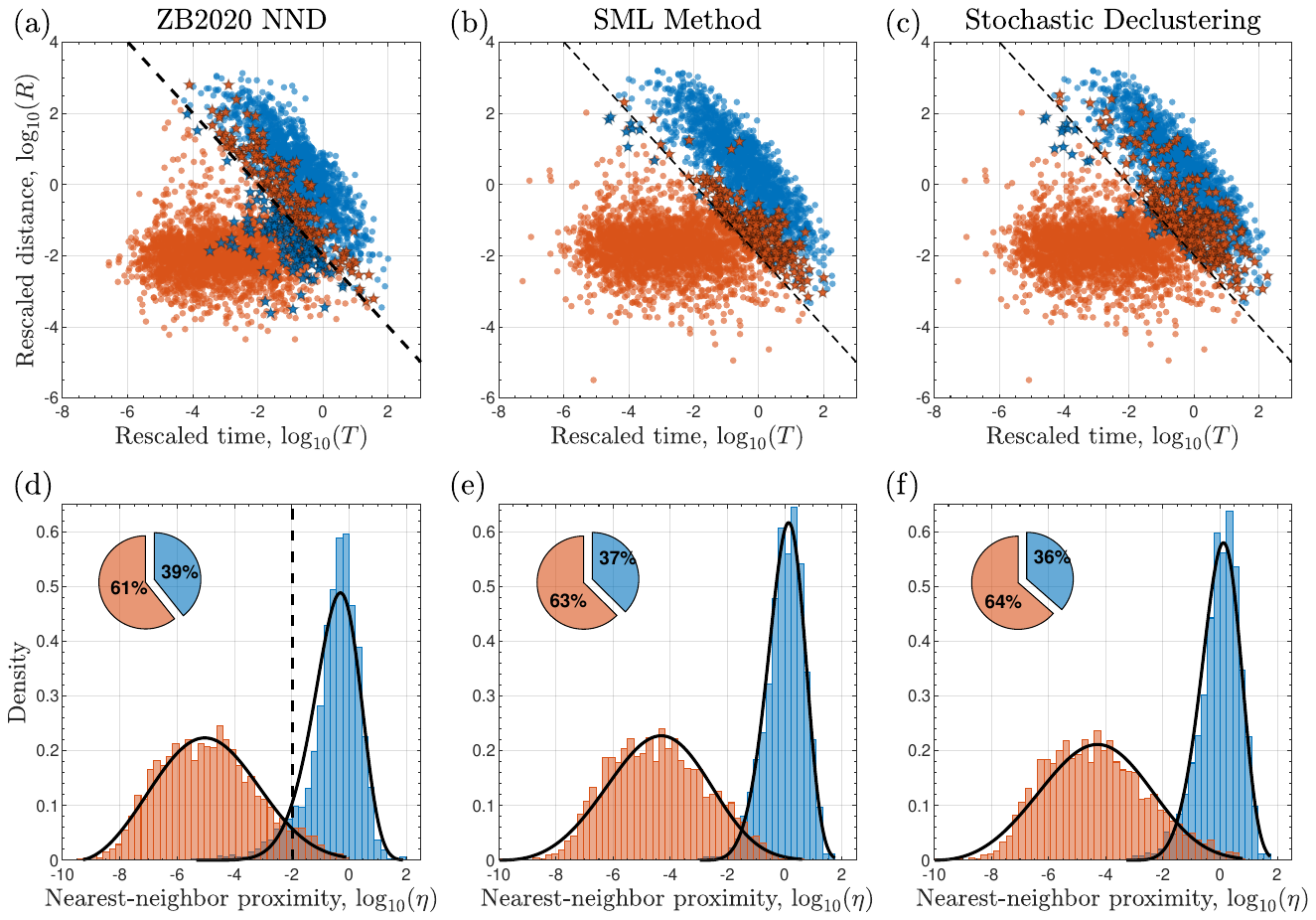}}

\caption{Declustering comparison of the three methods applied to the real seismic catalog for Italy. The symbols and all other features are the same as in Figure~\ref{fig4}.}
\label{fig5}
\end{figure}

The application of the three declustering methods to the real earthquake catalogs from Southern California and Italy are illustrated in Figures~\ref{fig4} and~\ref{fig5}, respectively. Clearly, all three algorithms perform comparably and make similar predictions for the bulk of the data. However, similar to the synthetic case, all three methods identify some aftershocks and background events beyond the NND threshold (Figure \ref{fig4}(a)-(c), blue and red stars). This correlates with the overlap in their respective $\eta$ distributions, Figure~\ref{fig4}(d)-(f). For both components in Figure \ref{fig4}(a)-(c), the outlier events identified beyond the threshold appear to conform to their respective modal shapes. The majority of additional aftershocks identified lie within a horizontally-oriented ellipse while the additional background events mostly trend along the downward diagonal. Other studies have demonstrated that these shapes are consistent with the nature of tectonic seismicity. \cite{ZaliapinB13a} noted that the non-homogeneous Poisson process represented in two-dimensional NND space results in an angled ellipse that trends along the downward diagonal at large $\eta$. Clustered activity typically groups in a subset of the NND space at smaller $\eta$, and the horizontal orientation is due in part to the inhomogeneous distribution of fault lines and in part to limits of seismic network location resolution.

\begin{figure}[!ht]
\centering
{\includegraphics*[width=0.46\textwidth, trim=17mm 66mm 18mm 65mm]{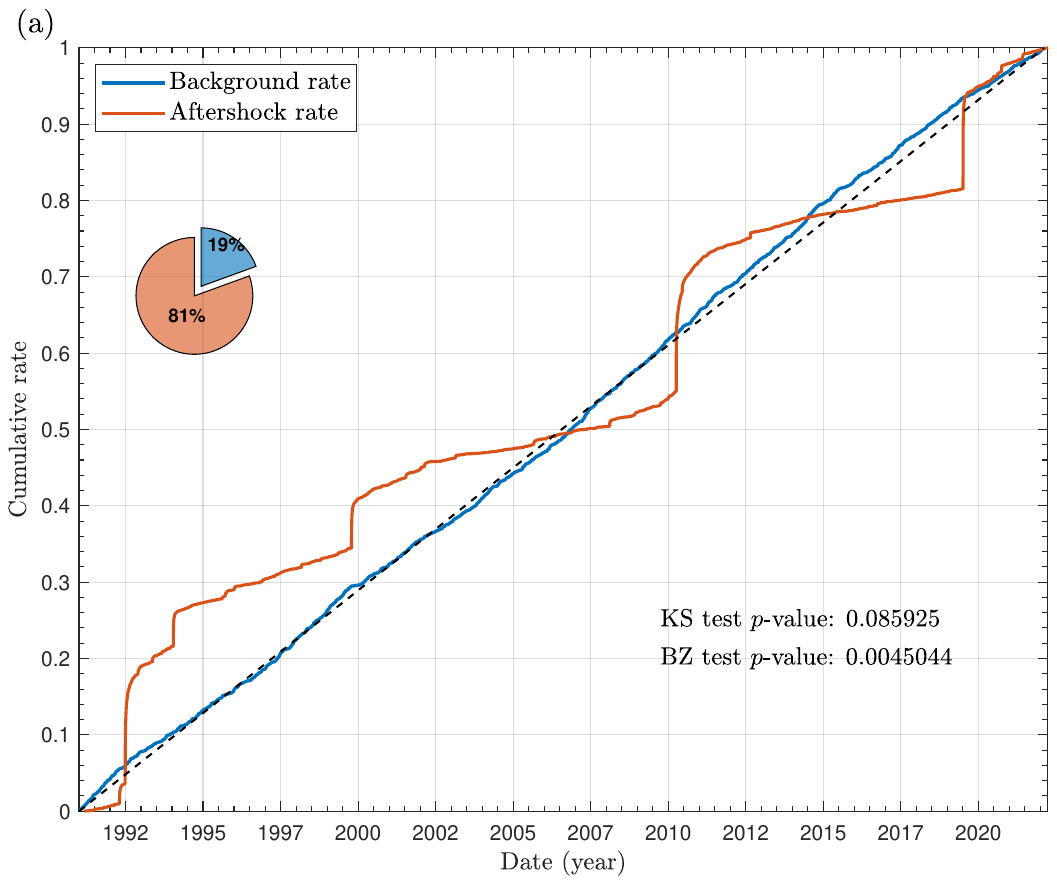}}
{\includegraphics*[width=0.46\textwidth, trim=17mm 66mm 18mm 65mm]{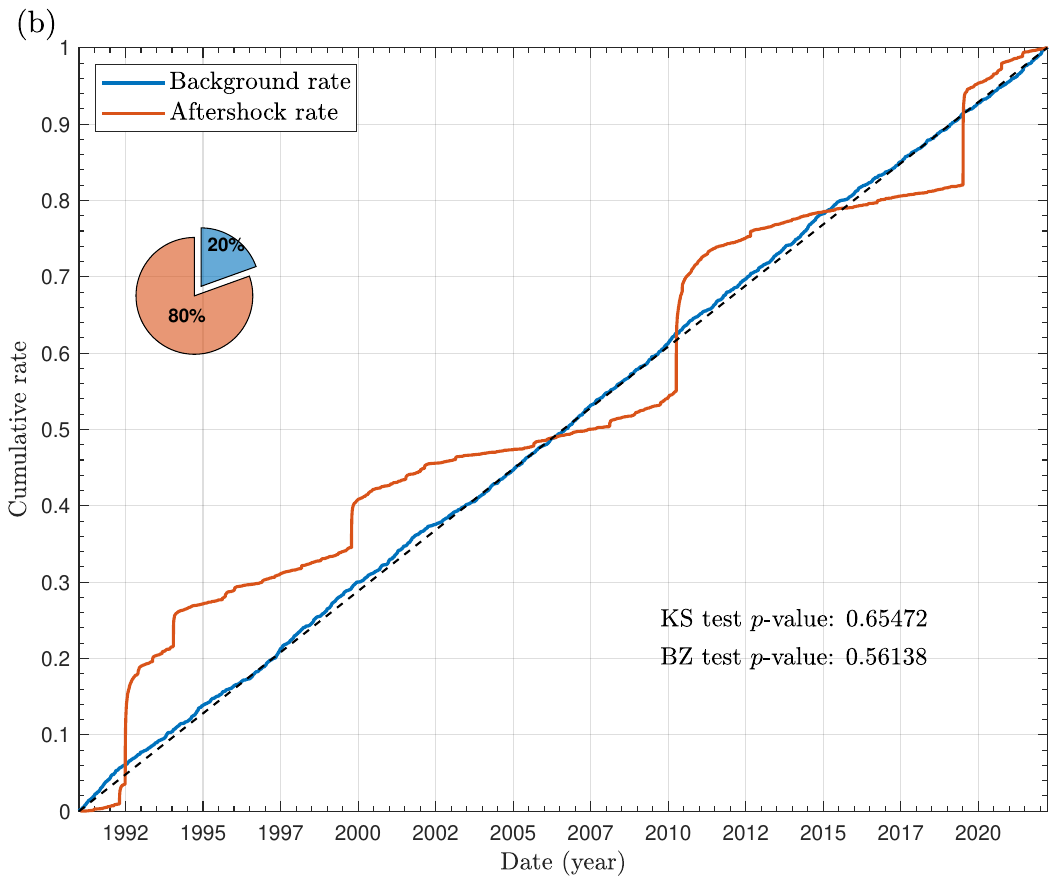}}

{\includegraphics*[width=0.46\textwidth, trim=17mm 66mm 18mm 65mm]{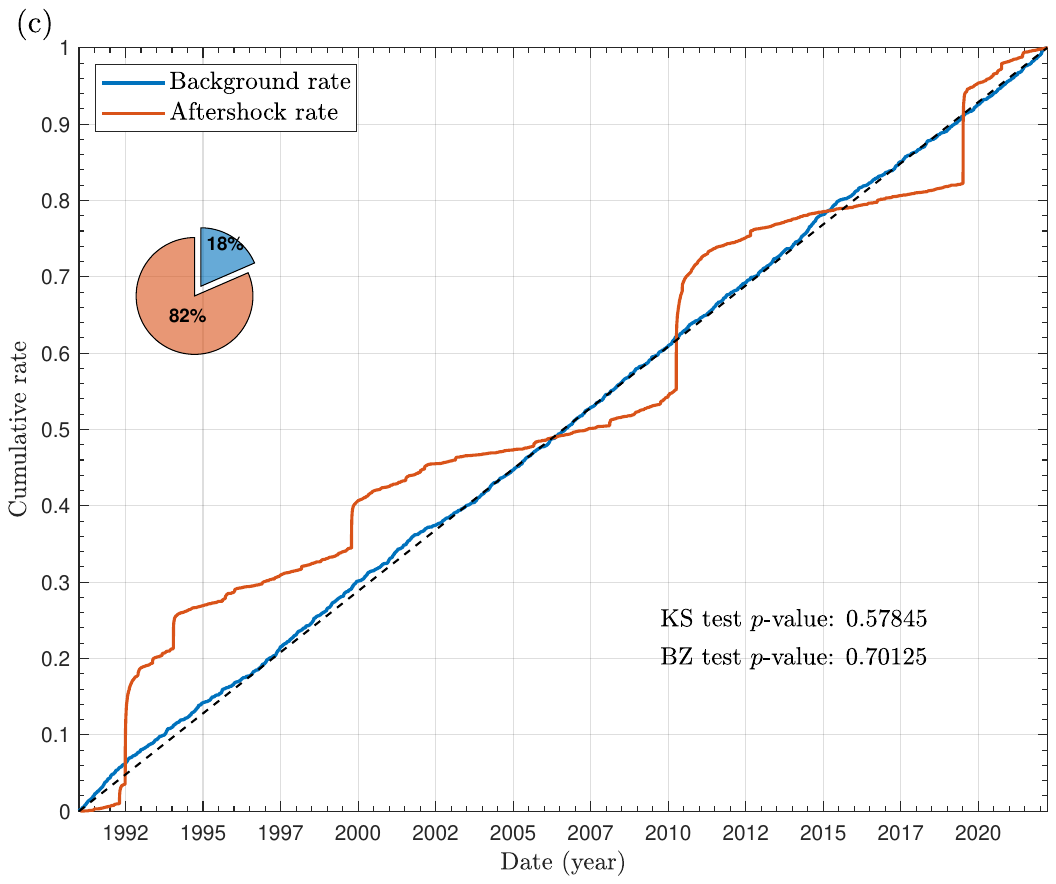}}

\caption{Earthquake declustering comparison for the Southern California catalog using (a) the NND method by \cite{ZaliapinB20a}, (b) the SML method, and (c) the SD method by \cite{ZhuangCOC05a}. Normalized cumulative seismic rates for background events and aftershocks are plotted in each case. The results of the three statistical tests are reported as the corresponding $p$ values.}
\label{fig6}
\end{figure}

The normalized cumulative rates of background events and aftershocks for Southern California are plotted in Figure~\ref{fig6}. The declustering was performed using the method of \cite{ZaliapinB20a} (Figure~\ref{fig6}a); the SML method developed in this study (Figure~\ref{fig6}b); and the SD method by \cite{ZhuangOV02a} (Figure~\ref{fig6}c). We also applied two statistical tests (the KS, and BZ tests) to the background events to test the null hypothesis that these events followed the homogeneous Poisson process. The results of the tests are reported as the corresponding $p$ values in Table~\ref{tab3}. The values of $p$ less than $0.05$ reject the Poisson null hypothesis. The corresponding results of the declustering of seismicity in Italy are given in Figure~S4.

\begin{table}
\tbl{The results, $p$ values, of the statistical tests for Southern California and Italy using the three declustering methods.\label{tab3}}
{\begin{tabular*}{.6\columnwidth}{@{\extracolsep\fill}lcccccc}
   \multicolumn{1}{c}{} & \multicolumn{2}{c}{\textbf{Southern California}} & \multicolumn{2}{c}{\textbf{Italy}} \trowsep 
\toprule
   \textbf{Test}      & KS               & BZ                & KS      & BZ        \trowsep
\toprule
   \textbf{Method}    &                  &                   &         &           \trowsep
   \hspace{2mm}ZB2020 & $\mathbf{0.086}$ & $0.0045$          & $5.3\mathrm{e-}5$ & $0.002$  \trowsep
   \hspace{2mm}SML    & $\mathbf{0.655}$ & $\mathbf{0.561}$  & $0.019$ & $0.040$   \trowsep
   \hspace{2mm}SD     & $\mathbf{0.578}$ & $\mathbf{0.701}$  & $0.005$ & $0.028$   \trowsep
\hline\hline
\end{tabular*}}
{}
\end{table}

\begin{figure}[!ht]
\centering
{\includegraphics*[width=.8\textwidth, scale=0.5, viewport= 15mm 45mm 200mm 240mm]{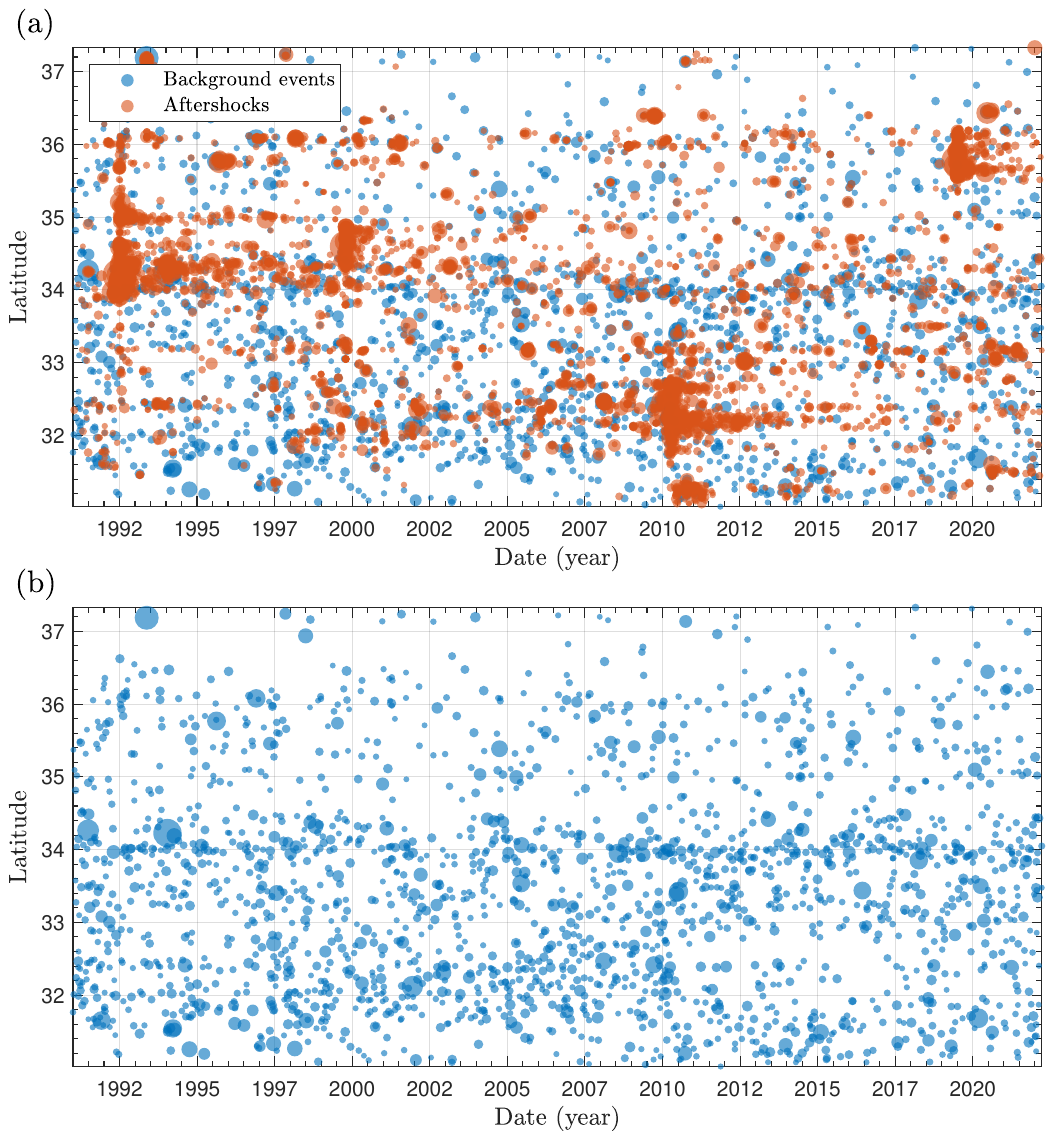}}

\caption{Comparison of the full and declustered catalogs for Southern California using the SML method. (a) The full catalog of events starting from 1991/01/01 to 2022/03/31 with varying latitudes is plotted. Background events are shown as solid blue circles and the aftershocks are shown as solid orange circles. (b) The declustered catalog with aftershocks removed is plotted.}
\label{fig7}
\end{figure}

To visualize the full and declustered catalogs for Southern California, we plot the evolution of the seismicity for the range of event latitudes and during the study target time interval (Figure~\ref{fig7}). In Figure~\ref{fig7}(a) one can see all the events above magnitude $m\ge3.0$, where aftershocks are marked with orange color and background events with blue color. In Figure~\ref{fig7}(b) only background events are plotted. The same plots for Italian seismicity are given in Figure~S5.

\begin{table}
\tbl{The computed $b$-values, of the GR relationship for background events and aftershocks for Southern California and Italy using the three declustering methods.\label{tab4}}
{\begin{tabular*}{.7\columnwidth}{@{\extracolsep\fill}lcccccc}
   \multicolumn{1}{c}{} & \multicolumn{2}{c}{\textbf{Southern California}} & \multicolumn{2}{c}{\textbf{Italy}} \trowsep 
\toprule
                      & $b$ (background) & $b$ (aftershocks) & $b$ (background) & $b$ (aftershocks) \trowsep
\toprule
   \textbf{Method}    &                  &                   &                  &                \trowsep
   \hspace{2mm}ZB2020 & $0.80\pm0.04$    & $1.04\pm0.02$     & $1.02\pm0.05$    & $1.19\pm0.05$  \trowsep
   \hspace{2mm}SML    & $1.04\pm0.05$    & $0.97\pm0.02$     & $1.13\pm0.05$    & $1.11\pm0.04$  \trowsep
   \hspace{2mm}SD     & $1.06\pm0.05$    & $0.97\pm0.02$     & $1.12\pm0.06$    & $1.11\pm0.04$  \trowsep
\hline\hline
\end{tabular*}}
{}
\end{table}

\begin{figure}[!ht]
\centering
{\includegraphics*[width=.8\textwidth, scale=0.6, viewport= 15mm 35mm 195mm 245mm]{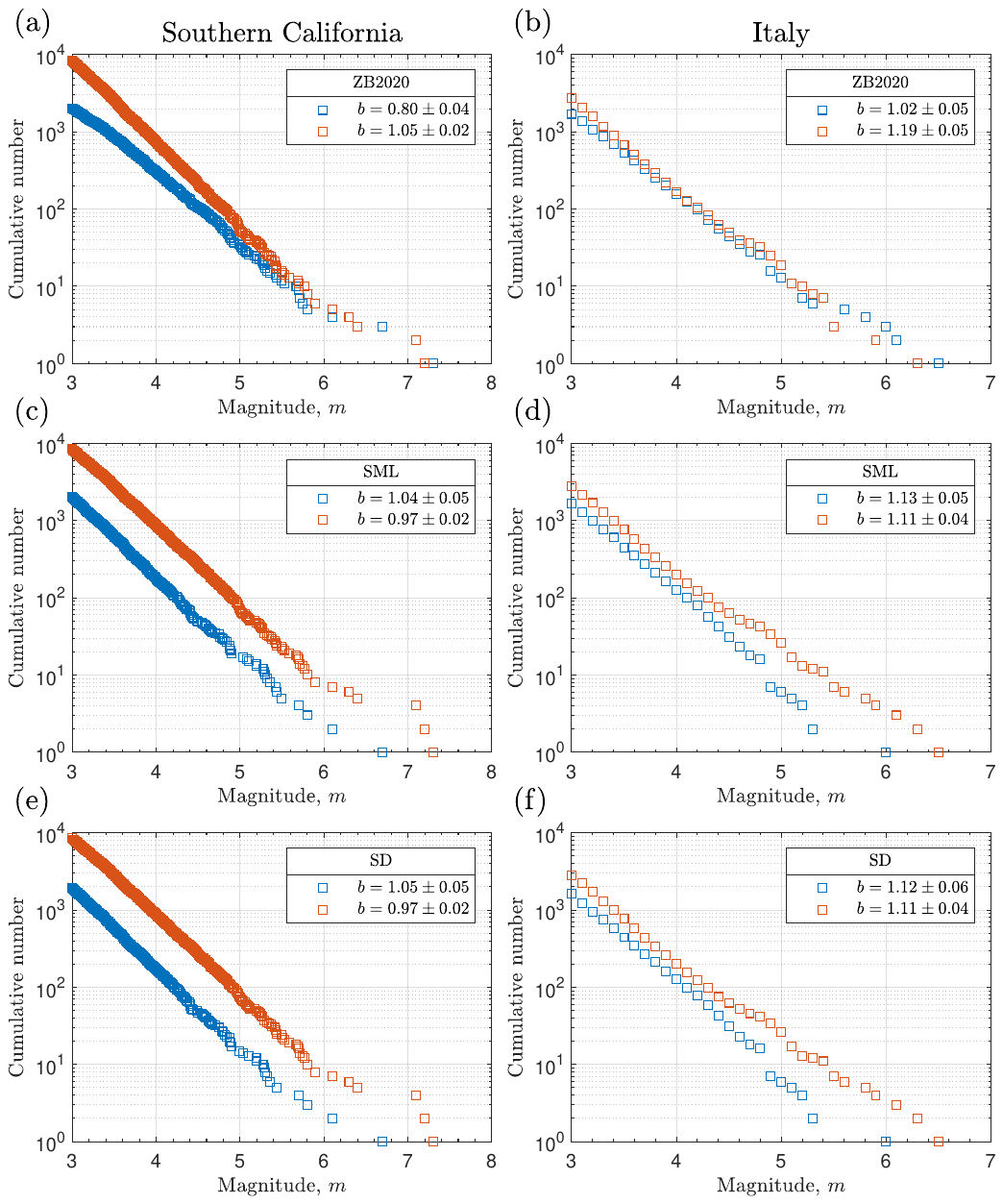}}

\caption{Comparison of the frequency-magnitude statistics and the estimated GR $b$-values for the declustered catalogs in Southern California and Italy. The declustering was performed using (a,b) the ZB2020 method; (c,d) the SML method; and (e,f) the SD method. Blue symbols are the cumulative numbers of background events and red symbols are the cumulative numbers of aftershocks.}
\label{fig8}
\end{figure}

We also computed the frequency magnitude distributions for the background and clustered events for seismicity in Southern California and Italy and fitted the GR relation to estimate the corresponding $b$-values. This was done for each declustering method. The results are shown in Figure~\ref{fig8} and the $b$-values are reported in Table~\ref{tab4}. The SML and SD methods produced similar results with comparable $b$-values between background events and aftershocks. In contrast, the ZB2020 method produced $b$-values that were significantly different for aftershocks and background events.

\section{Discussion}\label{discussion}

To asses the performance of the SML algorithm, we compared its output with two other declustering algorithms, i.e. the NND-based one (ZB2020) by \cite{ZaliapinB20a} and the SD method by \cite{ZhuangOV02a}. These three declustering algorithms were applied to real and synthetic seismicity in Southern California and Italy. For the synthetic seismic catalogs, the events were labeled as background events or aftershocks during the simulations as the branching algorithm was used to simulate the ETAS model. Therefore, it was straightforward to apply all the three declustering algorithms to these catalogs and compare with the ground truth. This is illustrated in Figure~\ref{fig3} for synthetic California seismicity and in Figure~S3 for Italy. Overall, the SML algorithm outperformed the other two algorithms. Both the ZB2020 and SD algorithms employ the random thinning procedure that can result in misclassification of both background and aftershock events. This is illustrated as yellow and purple colored points in Figure~\ref{fig3}(a,b). A yellow color indicates true aftershock events that were classified as background events by each declustering algorithm, Figure~\ref{fig3}(g)-(i). A purple color indicates true background events that were classified as aftershocks. The overall accuracy of the SML exceeded the other two declustering algorithms while more accurately separating background events from aftershocks. At the same time, all the three algorithms produced a comparable proportion between background and clustered events.

The SML method significantly outperformed the ZB2020 method when declustering the synthetic catalogs for Southern California, Figure~\ref{fig3} (Figure~S3 for Italy) in terms of overall accuracy and better identifying the background events (89.3\% versus 65.9\% for ZB2020 method). The ZB2020 method depends critically on the parameter $\alpha_0$ defined in \cite{ZaliapinB20a}. By varying this parameter one can significantly shift the proportion of identified background and aftershock events. However, it does not significantly improve the accuracy of the method. In our analysis, we used the value of $\alpha_0=0.7$ for Southern California and $\alpha_0 = 1.2$ for Italy. We also used a non-zero factor $w=\frac{1}{2}b$ in the definition of the exponential weight $10^{-wm_i}$ for the rescaled distances (\ref{etaij}) in contrast to the analysis performed in \cite{ZaliapinB20a}.

When the SML method was applied to two real earthquake catalogs, the analysis revealed that the method produced a more realistic declustering of background events by identifying a more clustered cloud of aftershocks shown in Figure~\ref{fig4} for Southern California and in Figure~\ref{fig5} for Italy. The dashed black lines indicate the separation of events based on the standard NND approach at a given $\eta_0$ threshold. All three methods identified the additional background events (blue stars) below this line and additional aftershocks (red stars) above this line. However, the ZB2020 and SD declustering methods identified more aftershock further away in rescaled time and distance compared to the SML method. In contrast, the original NND approach \citep{ZaliapinGKW08a,ZaliapinB13a} falls short in effectively discriminating events near its threshold, where no modal overlap is accounted for and an accuracy trade-off occurs wherever it is placed. In Figure \ref{fig3}(a)-(c) for synthetic seismicity and in Figures~\ref{fig4} and \ref{fig5} for real seismicity, we see that prediction errors clearly stack along the threshold for the traditional NND approach. If we were to shift the dividing line towards larger $\eta$ (i.e. towards the upper-right of the joint histograms), we would simultaneously reduce the aftershock prediction error and increase the background error, and vice versa. On the other hand, the ZB2020, SML and SD methods are not impeded by this fixed threshold and can ''cross over'' to select outlier events that lie far away from their group. In Figure \ref{fig3}, we observe that, by avoiding the threshold-based limitation, the three methods can accommodate overlapping background and aftershock distributions. This is also more physical, as one would expect both distributions to tail off and decay as opposed to terminate at a fixed point. 

For declustering of real catalogs, in contrast to the synthetic ones, we don't know the ground truth. As a result, one can perform additional statistical tests for independence or stationarity of declustered background events. Here, we implemented and applied two statistical tests to check the null hypothesis that the declustered background events follow a homogeneous Poisson process in time \citep{LuenS2012a,ZaliapinB20a}. The results show that the declustered catalogs for Southern California pass at least one test among the two applied for each method (Table~\ref{tab3}). This is also illustrated in Figure~\ref{fig6} for the normalized cumulative plot of background and clustered events, where declustered events follow very closely the straight diagonal line. This indicates that the background rate remains relatively constant in time. The obtained results are also consistent with the study of \cite{MarsanL08a}, where the authors applied the stochastic declustering approach to seismicity in Southern California. The obtained proportion of background and aftershock events is also consistent with our findings. It was reported that 19.5\% of events constituted the background seismicity \cite{MarsanL08a}.

The analysis of the seismicity in Italy does not show a similar behavior compared to Southern California (Figure~S4). There is a consistent deviation of the normalized cumulative rate of the declustered events for all three methods. As a result, none of the declustered catalogs pass the two statistical tests and the Poisson hypothesis is rejected at a 0.05 significance level (Table~\ref{tab3}). This signifies that the temporal background rate for seismicity in Italy does not follow a stationarity assumption. One possible explanation for this is the complicated geological setting of the region that is delineated by the boundary given in Figure~\ref{fig1}(b). As a result, the temporal background rates vary across the region and this results in the deviation of the total normalized cumulative rate from the straight line (Figure~S4). 

Several studies have shown that background seismicity is well-represented by a time-stationary, space-inhomogeneous Poisson process whose inter-event times follow the exponential distribution. This assumption is also used in the standard PSHA applications. \cite{BaylissNM2019a} demonstrated that the one-dimensional $\eta$ distribution of nearest-neighbour distances is best fit by a mixture of two Weibull distributions. The black curves in Figures~\ref{fig4}(d)-(f) and \ref{fig5}(d)-(f) are fits of the Weibull probability densities for each subpopulation of $\eta$ distributions. For the subpopulations selected by the SML and SD methods, the Weibull distribution fits well the background components, particularly in case of the synthetic seismicity (Figure~\ref{fig3}(d)-(f) and Figure~S3). On the other hand, the distribution of $\eta$ for clustered subpopulations deviate from the Weibull distribution. This is particularly pronounced for the synthetic catalogs. The distribution for $\eta$ is effectively a product of two distributions for interevents in temporal and spatial scales. It was shown by \cite{ShcherbakovYTR05a} that the slope of the distribution of interevent times in the case of aftershocks is controlled by the Omori-Utsu law exponent $p$. Therefore, the observed power-law slope in the distribution of $\eta$ for the clustered events can be the result of the contribution of the power-law decay in the rate of events in time and space that is observed for aftershocks. This is less pronounced in case of the real earthquakes in Southern California and Italy, Figures~\ref{fig4}(d)-(f) and \ref{fig5}(d)-(f).

Our study can be compared with a similar approach to declustering undertaken by \cite{Aden-AntoniowFS2022a}. In contrast to that study, we introduced more learning features by considering next nearest-neighbors when using the NND metrics. We also considered the magnitude differences for nearest-neighbor events. The parameters of the ETAS model were also estimated from a given seismogenic region to reflect its specific conditions. We also performed testing of the produced declustering catalogs to analyze their point process properties. A significant limitation of \cite{Aden-AntoniowFS2022a} method was the bias-variance trade-off, stemming from the random ETAS parameters used to generate training data. The motivation behind this was to expose the model to different aftershock distributions, thereby reducing prediction variance (''overfitting'') and improving generalizability. Unfortunately, this approach simultaneously serves to increase model bias. As a qualitative example, consider two training catalogs, $T_1$ and $T_2$, both generated using identical ETAS parameters with the exception of the spatial decay parameter $\gamma$, where $\gamma_1 \gg \gamma_2$. The first catalog would teach the classifier to correlate magnitude difference closely with inter-event distance between nearest-neighbours when predicting aftershock behaviour. The second catalog would then force the algorithm to unlearn this association. Because the classifier itself is blind to the parameters, its only course of action would be to reduce feature importance for these cases (i.e. increasing bias or ''underfitting''). Further, more complex counter-learning outcomes would arise from additional changes to other ETAS parameters within the training set. We address this limitation, as well as some others, by attaining maximum-likelihood estimates of the ETAS parameters for each target catalog, which we then use to generate our synthetic training catalogs. Clearly, this introduces an expensive additional step to our approach. However our goal was to produce an ML-driven declustering algorithm with the best possible accuracy for practical use.

The proposed SML method can also be extended into a full 3D space by incorporating the depth information from each event. However, to train the SML algorithm one needs to use an appropriate 3D point process model to capture the realistic depth variation of seismicity \citep{GuoZZ2024a}. The NND method and feature computation can also be formulated in 3D by considering the hypocentral distance between events. Besides addressing the problem of declustering of earthquake catalogs, the 3D analysis can be used in illuminating the subsurface structural features that control the occurrence of earthquakes \citep{PiegariHM2022a}.

\section{Conclusions}\label{conclusions}

Earthquake declustering is a challenging task in statistical seismology and probabilistic seismic analysis. There is no a robust discriminating measure that can be used to separate earthquakes in a seismicity catalog into different classes without making certain assumptions about the earthquake generation process. However, it is commonly recognized that catalogs contain independent and triggered events, which form clusters in space and time. For example, seismicity on a regional scale in highly active seismic zones is dominated by aftershock sequences. In addition, earthquakes can form swarms and/or can be induced by various external forcing mechanisms.

In this work, we implemented an earthquake declustering algorithm based on the supervised machine learning approach. The developed SML declustering method does not require any tuning of model parameters. However, the training of the algorithm is based on a specific stochastic process to generated labeled synthetic seismicity. For this purpose, we use the spatio-temporal ETAS model. The SML algorithm is trained on specific features that are extracted from the earthquake catalog. For this, we incorporated the nearest-neighbor distance analysis approach to compute such features. After training the algorithm on synthetic seismicity associated with a specific seismogenic zone it can be applied to a real earthquake catalog to decluster it.

We have applied the SML declustering algorithm to seismicity in Southern California and Italy. The obtained results show that in Southern California, the catalog contains approximately 20\% background events and 80\% clustered aftershocks. For Italy, this proportion is 37\% for background events and 63\% for clustered events. The analysis of the frequency-magnitude statistics of the background and clustered events demonstrates that both types of events have comparable $b$-values and the declustering does not significantly affect this proportion. The applied statistical tests indicate that the declustered seismicity in Southern California follows a homogeneous Poisson statistics. On the other hand, the seismicity in Italy appears to deviate from the Poisson assumption. We speculate that this might be due to a more complex tectonic setting compared to Southern California and the mixture of several tectonic regimes that produces distinct background seismicity rates. The developed approach can be used in PSHA studies that employ the declustering of seismicity catalogs, in order to provide a more accurate depiction of regional, background seismicity and improve earthquake risk assessment.

\begin{datres}

The earthquake catalogs were downloaded from \url{https://scedc.caltech.edu/data/alt-2011-dd-hauksson-yang-shearer.html} \citep{HaukssonYS12a} for Southern California and from \url{https://terremoti.ingv.it/en/iside} \citep{ISIDe} for Italy. Last accessed in October 2024.

The Matlab software codes, that were developed to perform the analysis, can be downloaded from (\url{https://github.com/rshcherb/SMLdecluster}).

Supplemental Material provides additional results in Figures S1-S5.

\end{datres}

\section{Declaration of Competing Interests}

The authors acknowledge that there are no conflicts of interest recorded.\footnote{The authors acknowledge that there are no conflicts of interest recorded.}

\begin{ack}
This research has been supported by the NSERC Discovery grant. Useful and constructive comments from one anonymous reviewer and Matteo Taroni helped to improve the presentation and clarify the results.
\end{ack}

\vspace{20mm}
\noindent Robert Shcherbakov$^{1,2\ddag}$ and Sidhanth Kothari$^{1}$ \\
\\
$^{1}$Department of Earth Sciences, Western University, London, Ontario, \emph{N6A 5B7}, Canada\\
$^{2}$Department of Physics and Astronomy, Western University, London, Ontario, \emph{N6A 3K7}, Canada\\
\\[5mm]
$^\ddag$E-mail: rshcherb@uwo.ca

\end{document}